\documentclass[usenatbib]{mn2e}
\usepackage{natbibmnfix,graphicx,times}

\newcommand{\bq}{\begin{equation}}
\newcommand{\eq}{\end{equation}}
\newcommand{\bqa}{\begin{eqnarray}}
\newcommand{\eqa}{\end{eqnarray}}
\newcommand{\deriv}{{\rm d}}
\def\fun#1#2{\lower3.6pt\vbox{\baselineskip0pt\lineskip.9pt
  \ialign{$\mathsurround=0pt#1\hfil##\hfil$\crcr#2\crcr\sim\crcr}}}
\def\lesssim{\mathrel{\mathpalette\fun <}}
\def\gtrsim{\mathrel{\mathpalette\fun >}}
\def\VEV#1{\left\langle #1\right\rangle}
\def\deltaptwo{{\VEV{\delta_p^2}}}

\title[Pair correlations and merger bias]{Pair correlations and merger bias}

\author[S.~R.~Furlanetto \& M.~Kamionkowski]{ Steven~R.
  Furlanetto\thanks{Email:sfurlane@tapir.caltech.edu} and Marc
  Kamionkowski\thanks{Email:kamion@tapir.caltech.edu}
  \\ 
California Institute of Technology, Mail Code 130-33, Pasadena,
CA 91125, USA}

\begin{document}

\maketitle

\begin{abstract}
We study analytically the possibility that mergers of haloes are more
highly clustered than the general population of haloes of comparable
masses.  We begin by investigating predictions for merger bias within
the extended Press-Schechter formalism and discuss the limitations
and ambiguities of this approach.  We then postulate that mergers
occur whenever two objects form within a (small) fixed distance of
each other.  We therefore study the clustering of pairs of points for
a highly biased population in the linear regime, for the overall mass
distribution in the quasilinear regime, and (using the halo model of
clustering) in the nonlinear regime.  Biasing, quasilinear evolution,
and nonlinear clustering all lead to nonzero reduced (or connected)
three-point and four-point correlation functions.  These higher-order
correlation functions can in many cases enhance the clustering of
close pairs of points relative to the clustering of individual points.
If close pairs are likely to merge, then the clustering of mergers may
be enhanced.  We discuss implications for the observed clustering of luminous $z=3$ galaxies  and for correlations of active galactic nuclei and galaxy clusters.
\end{abstract}


\section{Introduction}
\label{intro}

Galaxy clustering can be a useful tool to study the origin of
large-scale structure and to delineate the formation mechanisms of
various types of galaxies.  For example, it is now well appreciated
that objects forming from rare high-density peaks in the primordial
density distribution, such as bright galaxies at high redshifts or
galaxy clusters today, should be ``biased'' (i.e., more highly
clustered) relative to the more common lower-mass objects that more
closely trace the total-mass distribution \citep{Kaiser}.

A currently unanswered question is whether the growth history of
haloes can affect their clustering properties.  Cosmological simulations give confusing results.  \citet{Kolatt99} argued that merger-driven starbursts at $z \sim 3$ occur in small haloes that lie near larger ones:  thus they are more highly clustered than typical objects of the same mass (see also \citealt{Wechsler01}).  The simulations of \citet{Gottlober02} showed different clustering at $z=0$ between objects that had experienced a major merger and those that had not. \citet{Kauffmann02} also found a weak enhancement in the cross-correlation between objects undergoing major mergers and the general population, but only at small scales.  On the other hand, \citet{percivaletal} found no evidence for excess merger bias at $z=0$, where recently-merged objects were identified as haloes in which at least 50 per cent of constituent particles were not in a progenitor of at least equal mass at a fixed earlier redshift.  \citet{Evan} agreed at $z=3$, but if they modified the criterion to include all haloes that grew by 20 per cent or more (implicitly including smooth infall), the rapidly-growing sample had a substantial excess bias, making their clustering comparable to that of haloes with three times more mass.  Most recently, \citet{Gaoetal} examined a high--dynamic-range N-body simulation at $z=0$.  They found the clustering of low-mass recently-merged objects to be suppressed relative to the average.  For
example, in their lowest-mass bin (with a mass $\approx 2\%$ of the
characteristic halo mass), the $20\%$ youngest and oldest haloes are
under- and over-biased by $\sim 40\%$, respectively.  On the other
hand, in agreement with \citet{percivaletal}, they found that the
clustering of more massive objects is nearly independent of their age.
The verdict is clearly not yet in: how can we reconcile these
disparate results?

The question is not just academic.  Clustering is often used to infer
information about the host halo mass of particular galaxy populations
(e.g., \citealt{MoFuk96,Adelberger98,Giavalisco98}).  The possibility
that clustering depends on the merger history---which obviously also
strongly affects observables such as the star-formation
history---would call such inferences into question.  One example is
the discrepancy between the masses ($\sim10^{12}\, {\rm M}_\odot$) of
Lyman-break galaxies (LBGs) inferred from their clustering
(\citealt{Coles98,Giavalisco01,Porciani02,Adelberger05})
and the dynamical masses ($\sim 10^{11}\, {\rm M}_\odot$) inferred from the
broadening of nebular emission lines and kinematics
(\citealt{Pettini01,Erb03}).  This claimed discrepancy may simply be the difference between the mass in the central regions and the total mass \citep{Erb03, Cooray05}, but \citet{Wechsler01} and \citet{Evan} have proposed that it may also point to ``merger bias" if LBGs are galaxies that have recently
undergone mergers.  The problem is even more extreme for submillimetre-selected galaxies at $z>2$:  their dynamics imply total masses $M \la 10^{12} \, {\rm M}_\odot$ while clustering implies $M \ga 10^{13} {\rm M}_\odot$ \citep{Blain04}.  

More generally, to what extent does clustering depend on factors other
than the halo mass?  Will selection techniques that trace recent
episodes of star formation (such as Lyman-break or Ly$\alpha$ line
selection) yield more highly clustered objects than techniques
sensitive to the total stellar mass (such as infrared observations),
even if the typical halo masses in the surveys are identical?  
Quasars and other active galactic nuclei (AGN) may also be triggered
by galaxy mergers.  Their clustering has been used to infer the
properties of the host galaxy \citep{LaFranca98,Adelberger05b} and of
the quasar (especially its lifetime;
\citealt{Haiman01,Martini01,Adelberger05a}).  How will the bias of
mergers (if it exists) affect such estimates?
Will recently-merged galaxy clusters trace the underlying mass
distribution differently than relaxed clusters?  All of these
questions have implications for our understanding of both galaxy
formation and the large-scale structure of the universe.

In this paper, we take an analytic approach that complements the
numerical studies and may aid in their interpretation.  We begin in
Section \ref{eps} by considering the question of ``merger bias''
within the context of the widely-used linear-bias model.  We show that existing techniques cannot adequately
answer this question, so we then go on to consider other approaches.
To be more precise, in Sections 3--7, we derive analytic results
for the clustering of
close pairs of galaxies in several clustering models.  We consider the
clustering of close pairs when galaxies Poisson sample (a)
the overall mass in a Gaussian random field; (b) the high-density
peaks in a primordial Gaussian random field; (c) the overall mass in
the quasilinear regime; and (d) the overall mass in the nonlinear
regime described by the halo-clustering model.  We find that the
clustering of close pairs of galaxies can be enhanced, sometimes
significantly, relative to the galaxies in many of these cases.  We
speculate that if close pairs are likely to merge, then a pair bias
will imply a merger bias, although we do not make this statement
precise.  If a pair bias does in fact lead to a merger bias, then our
results are consistent with a solution to the LBG puzzle.  We also
briefly discuss other observable implications of our results.

\section{A first look at merger bias}
\label{eps}

We will first attempt to compute the bias of merging objects via their
number densities and the ``peak-background split" approach to bias \citep{efstathiou88, ColeKaiser, Mo96}.  We define the number density $n_m\, \deriv m_1\, \deriv m_2$ of mergers between
haloes in the mass range $m_1 \rightarrow m_1+ \deriv m_1$ and those in the
mass range $m \rightarrow m_2 + \deriv m_2$ via
\bq
     n_m(m_1,m_2,z) = n(m_1,z)\,n(m_2,z) \, Q(m_1,m_2,z) \, \Delta t,
\label{eq:mker}
\eq
where $n(m,z)\, dm$ is comoving number density, at redshift $z$,
of haloes with masses $m \rightarrow m+\deriv m$ and $Q(m_1,m_2,z)$ is the
merger kernel with units of volume per unit time.  We take $\Delta t$
to be some finite time interval within which the mergers of interest
take place; note that we assume it to be sufficiently small that
the underlying halo populations do not evolve significantly.

To compute the bias, we simply need to know how each of these terms
varies (to linear order) with the mean density $\delta$ in some large patch.  
For example, the \citet{Press74} mass function is
\bq
n(m,z) = \sqrt{\frac{2}{\pi}} \, \frac{\bar{\rho}}{m^2} \,
\frac{\delta_c(z)}{\sigma} \, \left| \frac{ \deriv \ln \sigma}{\deriv \ln
  m} \right| \, \exp \left[ - \frac{\delta_c^2(z)}{2 \sigma^2} \right],
\label{eq:nps}
\eq 
where $\delta_c$ is the fractional-overdensity threshold for
spherical collapse,
$\bar{\rho}$ is the mean background density, and $\sigma^2$ is the
fractional-density variance smoothed on scale $m$.  Note that we follow the
convention in which $\sigma$ is independent of redshift, while
$\delta_c(z)$ is the (linear-extrapolated) density threshold at
redshift $z$.  This distribution can be derived in terms of a
diffusion problem in $(\sigma^2,\delta)$ space with an absorbing
barrier at $\delta=\delta_c$ \citep{Bond91}.  Such an approach makes
it obvious that the abundance of haloes in a region of (linear-extrapolated)
overdensity $\delta$ and mass $M$ (corresponding to $\sigma_M$) will take
the same form, but with a shift in the origin \citep{Lacey93}:
\bqa
n(m,z|\delta,M) & = & \sqrt{\frac{2}{\pi}} \, \frac{\bar{\rho}}{m^2} \,
\frac{\sigma^2[\delta_c(z) - \delta]}{(\sigma^2-\sigma_M^2)^{3/2}} \,
\left| \frac{ \deriv \ln \sigma}{\deriv \ln 
  m} \right| \nonumber \\
& & \times  \exp \left\{ - \frac{[\delta_c(z)-\delta]^2}{2
  (\sigma^2-\sigma_M^2)} \right\}.
\label{eq:npscond}
\eqa

To find the linear bias, \citet{Mo96} first take the large-scale
limit $M \rightarrow \infty$ (or $\sigma_M \rightarrow 0$).  The
overdensity of haloes in a region of physical volume $V$ is
\bq
\delta_h = \frac{n(m,z|\delta) \, V \, (1+\delta_z)}{n(m,z) V} - 1,
\label{eq:dh}
\eq
where $\delta_z$ is the true overdensity at redshift $z$ (without
linear extrapolation) and the $(1+\delta_z)$ factor in the numerator
accounts for the fact that an overdense region is larger in Lagrangian
space than in physical space.  Expanding equation (\ref{eq:npscond})
to linear order, we find 
\bqa 
\delta_h & \approx & \delta_z \left[1 +
\frac{\nu^2 - 1}{\delta_c(z=0)} \right] + {\mathcal O}(\delta_z^2) \\
& \equiv & b_h(m,z) \, \delta_z + {\mathcal O}(\delta_z^2), \nonumber
\label{eq:biasdefn}
\eqa
where we have let $\nu = \delta_c(z)/\sigma$.  This defines the
usual bias $b_h(m,z)$ for haloes of mass $m$ at redshift $z$.

\subsection{The extended Press-Schechter merger kernel}

To compute the merger bias, we need to perform a similar expansion on
the kernel $Q$. The usual model for this quantity comes from the
extended Press-Schechter merger rates of \citet{Lacey93}.
Unfortunately, as we will see explicitly below, this formalism is inherently unable to address our problem:  the large-scale bias of mergers disappears from the calculation.  Letting $S \equiv \sigma^2$, \citet{Lacey93} define $f(S_1,\delta_{c1}|
S_T,\delta_{cT})$ to be the fraction of excursion-set trajectories
that first cross $\delta_{c1} > \delta_{cT}$ at $S_1>S_T$, given that
they first cross $\delta_{cT}$ at $S_T$ (here the subscript $T$ refers to the total mass).  This is exactly equivalent
to $n(m,z|\delta,M)$ in equation~(\ref{eq:npscond}) with the
identifications $(S_1 \leftrightarrow m)$, $(S_T \leftrightarrow M)$,
$\delta_{c1}=\delta_c(z)$, and $\delta_{cT}=\delta$; the only
difference is that here we assume $M$ to be in a collapsed halo at a
later redshift.  To obtain the merger rate, we will need
$f(S_T,\delta_{cT}|S_1,\delta_{c1})$ instead: given a halo at some
early time, this function describes the distribution of objects to
which that halo can belong at some later time.  By Bayes' theorem,
it is simply
\bq
f(S_T,\delta_{cT}|S_1,\delta_{c1}) \, \deriv S_T =
f(S_1,\delta_{c1}|S_T,\delta_{cT}) \,
\frac{f(S_T,\delta_{cT})}{f(S_1,\delta_{c1})} \, \deriv S_T,
\label{eq:bayes}
\eq
where $f(S,\delta_c)$ is the unconditional first-crossing distribution
(i.e., the normal Press-Schechter halo mass function).  The extended
Press-Schechter formalism defines $\deriv^2 p/\deriv m \deriv t$, the
probability that a halo of mass $m_1$ will merge with an object of
mass $m_2 \equiv m_T - m_1$ within an infinitesimal time interval
$\deriv t$, from the limit of this distribution as $\delta_{cT}
\rightarrow \delta_{c1}$.  In other words, it is the probability that
the object will join a larger halo in the time interval of interest.
The total merger rate $n_m(m_1,m)$ is then this limit (transformed to
mass and time units) multiplied by $n(m_1)$.

For our problem, we need to know the dependence of each of these
quantities on the large-scale density $\delta_b$ (defined over some
mass with $S_b \ll S_1,S_T$).  The unconditional distributions are
easy: $f(S,\delta_c) \rightarrow f(S,\delta_c|S_b,\delta_b)$, just
like the conditional mass function.  We are thus left with the
progenitor distribution $f(S_1,\delta_{c1}|S_T,\delta_{cT})$ within
the large-scale region.  Recall, however, that this distribution
follows from a diffusion problem with origin $(S_T,\delta_{cT})$.  It
must therefore be independent of the behavior on scales $S_b < S_T$;
we only need to know that it passes $\delta_{cT}$ for the first time
at $S_T$ to compute the progenitor distribution.  
This step is obviously where the extended Press-Schechter formalism fails:  it is completely unable to incorporate the large scale environment of merger events, so it cannot make predictions about their bias.  
To see this explicitly, we calculate how merger densities vary with $\delta_b$:
\bqa
n_m(m_1,m|\delta_b) & \propto & n(m_1|\delta_b) \, \frac{\deriv^2
  p(\delta_b)}{\deriv m \, \deriv t} 
\, \Delta t \\
& \propto & f(S_1,\delta_{c1}|\delta_b) \times
\frac{f(S_T,\delta_{cT}|\delta_b)}{f(S_1,\delta_{c1}|\delta_b)} \\
& \propto & f(S_T,\delta_{cT}|\delta_b).
\label{eq:epsnm}
\eqa
Thus, according to the extended Press-Schechter model, $n_m$ varies with
density in precisely the same way as the number density of
haloes with the same final mass $m_T$.  Clearly there is \emph{no} merger
bias in this picture, but only because the formalism is unable to address the relevant question.

Thus the conclusion of this model is not one that we can trust.  In addition to this difficulty, there is the deeper one pointed out by \citet{Benson05}, who showed that the extended Press-Schechter merger
rates are mathematically self-inconsistent (calling into question the association of trajectory jumps with mergers).  While it has proven useful in a variety of contexts for galaxy formation, the extended Press-Schechter formalism is manifestly not appropriate for investigating merger bias.

\begin{figure}
\begin{center}
\resizebox{8cm}{!}{\includegraphics{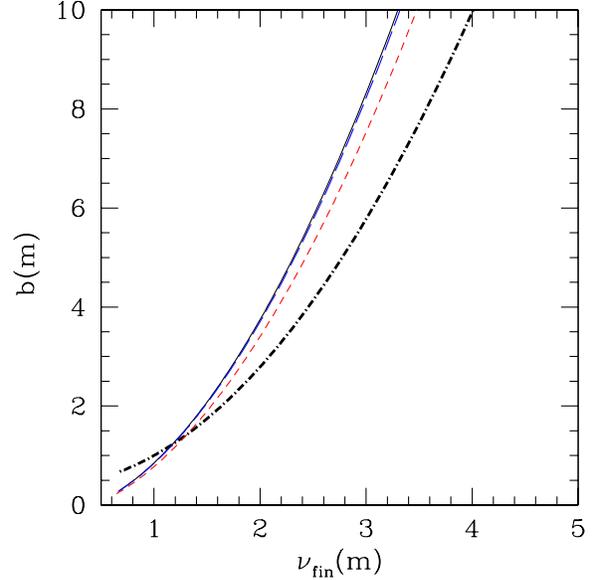}}\\%
\end{center}
\caption{Merger bias at $z=3$.  The dot-dashed line shows the normal
  halo bias $b_h$ for the final merger product.  The thin solid,
  long-dashed, and short-dashed curves take $m_2/m_1=1,\,0.5$, and
  $0.1$, respectively.}
\label{fig:bmz3}
\end{figure}

\subsection{A density-independent merger kernel}

Unfortunately, at this time, there are no fully developed alternatives
to the extended Press-Schechter formalism (but see \citealt{Benson05}
for first steps in this direction).  We therefore obviously cannot
compute the variation of $Q$ with the large-scale density.  Instead we
will consider the simplest possible model.  We will assume that the
merger kernel $Q$ is independent of environment in the Lagrangian
space to which the Press-Schechter formalism is native: that is, the
merger rate varies with the local density only through the Lagrangian
number density of haloes.  This would be appropriate if, for example,
all Gaussian peaks within a fixed comoving distance merged with each
other, and if we neglect extra correlations between neighboring haloes.  In other words, we treat each of the two haloes independently of the other; clearly, this is not completely correct, because the large-scale bias does not describe the small-scale correlations between haloes (e.g., \citealt{Scann02}).  We emphasize, then, that our model is not meant to be quantitatively accurate but only to illuminate the dependence of the merger bias on the halo abundances.  In this case, we define the overdensity of mergers via
\bq
\delta_m \equiv \frac{N_m(m_1,m_2,z|\delta)}{n_m(m_1,m_2,z) V} - 1,
\label{eq:mbiasdefn}
\eq
where $N_m$ is the number of mergers in this volume.  Clearly $N_m
\propto n(m_1|\delta) \, n(m_2|\delta) \, V \, (1 + \delta_z)$.
Expanding to linear order, we find a merger bias
\bq
b_m = 1 + \frac{\nu_1^2 + \nu_2^2 - 2}{\delta_c(z=0)},
\label{eq:bmsimple}
\eq
where $\nu_1 \equiv \nu(m_1)$, etc.  

For a given final mass $\nu$, we can then compute the bias of mergers
as a function of the mass ratio.  We show some results at $z=3$ in
Figure~\ref{fig:bmz3} as a function of $\nu_{\rm fin} \equiv \nu(m_1 +
m_2)$.  Interestingly, in this model, $b_m > b_h$ for $\nu \gg 1$:
mergers between massive objects tend to occur in denser regions than
an average halo of the final mass (or in other words, younger systems
are more biased than older systems).  The behavior reverses at small
masses: younger systems are less biased than average.  Qualitatively,
a dark-matter particle in a halo with $\nu \ll 1$ must be in a
low-density environment; small-mass objects that have just formed will
typically be in lower-density environments than an average halo of
this type.

\begin{figure}
\begin{center}
\resizebox{8cm}{!}{\includegraphics{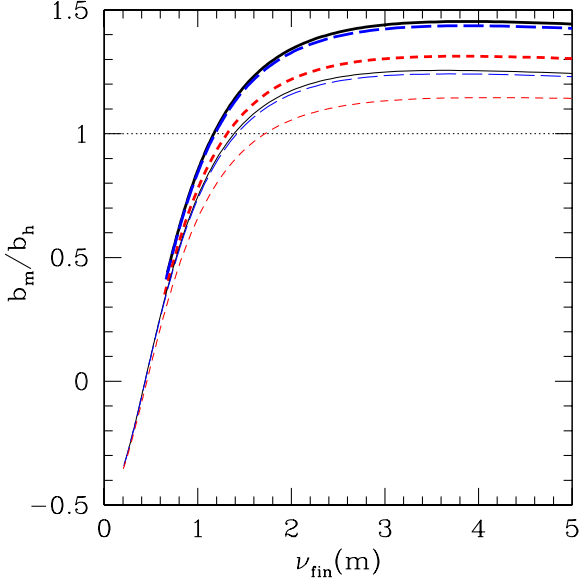}}\\%
\end{center}
\caption{Ratio between the merger bias $b_m$ and the halo bias $b_h$
  (of the final product).  The solid, long-dashed, and short-dashed
  curves take $m_2/m_1=1,\,0.5$, and $0.1$, respectively.  The upper
  thick and lower thin sets of curves take $z=3$ and $z=0$,
  respectively.}
\label{fig:bmzrat}
\end{figure}

Figure~\ref{fig:bmzrat} shows the ratio between the merger and halo
bias at both $z=3$ and $z=0$.  Note that it appears to asymptote to a
constant at large $\nu$.  This is simply $b_m/b_h \rightarrow (\nu_1^2 +
\nu_2^2)/\nu_{\rm fin}^2$; the excess bias will thus disappear when
one progenitor contains nearly all of the final mass.  Also, $b_m$ can
become negative for sufficiently small mass mergers: such events
preferentially occur in \emph{underdense} environments.  Note also
that in this model the merger bias at fixed $\nu_{\rm fin}$ depends on
redshift, even though the halo bias does not; this is because (for a
fixed mass ratio) the ratio $\nu_1/\nu_2$ does depend on redshift
through the scale dependence of the effective slope of the cold-dark-matter (CDM) power
spectrum.

Of course, it is not obvious that taking $Q$ to be constant in
Lagrangian space is the most reasonable assumption.  We could instead
have taken it to be independent of environment in physical
(Eulerian) coordinates.  Then the appropriate bias would be
\bq
b_m' = b_h(m_1) + b_h(m_2) = b_m + 1,
\label{eq:bmeul}
\eq
because in this case $Q \propto (1 + \delta_z)$ when expressed in the
Lagrangian space.  This would be appropriate if, for example, mergers
occurred only through random collisions in physical space.  In this
approximation, mergers are even more biased for large $\nu$ and less
antibiased for small $\nu$.  It is not clear which of these
assumptions is more physically plausible, but interestingly they both
predict positive bias ($b_m/b_h>1$) for mergers of massive haloes and
antibias ($b_m/b_h<1$) for mergers of sufficiently small haloes.

Comparison to the simulation results illuminates some of the
properties of $Q$ appropriate to halo growth.  \citet{Gaoetal} found
that, for small haloes at $z=0$, younger objects are less biased than
average.  This fits, at least qualitatively, with our $Q=\,$constant
results, which predict $b_m < b_h$ for $\nu \la 1$.  However, they
also found no evidence for age-dependent clustering in massive objects
(see also \citealt{percivaletal}).  This is in conflict with the
$Q=\,$constant results, which predict a $10$--$20\%$ enhancement to the
merger bias for large $\nu$.  Taken at face value, this implies that
the merger rate of massive objects must be \emph{suppressed} in dense
regions.  On the other hand, \citet{Evan} claimed a positive merger
bias for massive haloes in simulations at $z=3$.  The $Q=\,$constant
model provides an important clue that may explain this apparent
redshift evolution: it does indeed predict a larger merger bias at
early times.  The reason is that the merger bias depends on $\nu(m_1)
+ \nu(m_2)$ and not simply $\nu(m_1 + m_2)$.  The characteristic scale
of the mass function grows with time; because the CDM power spectrum
is not a simple power law, the relation between these three quantities
changes with time.  Thus, although the halo bias at a fixed $\nu$ is
independent of redshift, the bias of major mergers need not be.

\section{Clustering of Pairs}
\label{ncol}

The last Section showed that, until we have a self-consistent merger
kernel $Q$ that correctly incorporates the density dependence of the merger
rates, we cannot properly predict the linear merger bias within the
Press-Schechter model.  It is therefore worth considering other
approaches to merger bias to see what light they can shed.  In this
and the following Sections, we will examine a picture in which mergers
simply correspond to closely spaced objects.  Intuitively, such pairs
may merge because of (for example) nonlinear gravitational collapse
that brings objects closer together.  We will consider how close pairs
are biased relative to the objects themselves and show that, in
general, the pair bias differs from the halo bias.

Consider a population of galaxies with mean spatial density $n$.  Then
the differential probability to find a galaxy in an infinitesimal volume
element $\deriv V$ is $\deriv P=n\, \deriv V$.  The differential
probability to find one galaxy in $\deriv V_1$ centered on a position
$\vec r_1$ and another in $\deriv V_2$ centered on $\vec r_2$ is
$\deriv P = n^2 \, \deriv V_1 \, \deriv V_2 \, \left[ 1+ \xi(|\vec r_1
-\vec r_2|) \right]$, where $\xi(r)$ is the galaxy-galaxy
autocorrelation function.  The correlation function is the excess
probability, over random, to find two galaxies in differential volume
elements separated by a distance $r$.

There can never be more than one galaxy in an infinitesimal
volume element $\deriv V$.  However, we will
soon deal with close pairs of galaxies.  We will thus want to
know the probability to find {\it two} galaxies in one small,
but finite, volume element $\delta V$.  To be precise, we
take this volume element to be a sphere of radius $r_p$; then
$\delta V = (4\pi/3)r_p^3$.  The desired probability is then
\begin{eqnarray}
     \delta P &=& n^2 \int_{\delta V}\deriv^3 r_1\,\int_{\delta V}\deriv^3
     r_2\, \left[1+ \xi(|\vec r_1 -\vec r_2|) \right] \nonumber
     \\
     &\equiv& n^2 (\delta V)^2 \left(1+\deltaptwo \right),
\label{eqn:twogalaxies}
\end{eqnarray}
where
\begin{eqnarray}
     \deltaptwo &\equiv& (\delta V)^{-2} \VEV{
     \left[\int_{\delta V} \deriv^3 r \,\delta(\vec
     r - \vec x) \right]^2} \nonumber \\
     &=& (\delta V)^{-2}\int_{\delta_V} \deriv^3 r_1
     \int_{\delta V} \deriv^3 r_2
     \VEV{\delta(\vec r_1-\vec x)\delta(\vec r_2 -\vec x)}
     \nonumber \\
     &=& (\delta V)^{-2} \int_{\delta V} \deriv^3 r_1 \int_{\delta V}
     \deriv^3 r_2 \xi(|\vec r_1 -\vec r_2|),
\end{eqnarray}
is the variance of the density perturbation smoothed over a
spherical top hat of radius $r_p$.
If the correlation function can be approximated by a
power law, $\xi(r) \propto r^{-\alpha}$, for $r<r_p$, then
\begin{eqnarray}
     \VEV{\delta_p^2} &=& \frac{9}{2} \xi(r_p) \int_0^1\,
     x_1^2\, \deriv x_1\,
     \int_0^1\, x_2^2\, \deriv x_2\nonumber \\
     & &\times \int_{-1}^1\, \deriv \mu\,
     \frac{1}{\left(x_1^2 +x_2^2-2 x_1 x_2 \mu
     \right)^{\alpha/2}}.
\end{eqnarray}
For $\alpha=0$, the integral evaluates to $2/9$.  And for
$\alpha=1$, 2, and 3, it evaluates to 0.27, 0.50, and 5.0,
respectively.  The integral is 0.41 for $\alpha=1.8$.  

To begin, we take the radius of the sphere so that the probability to
find three or more galaxies is small compared with that to find
two.  Roughly speaking (neglecting corrections from higher-order
clustering that will become apparent below), this requires the
probability to find two galaxies in $\delta V$ to be small
compared with that to find one.  We thus require the radius
$r_p$ to be chosen small enough so that $n\, \delta V
\left(1+\deltaptwo\right) \lesssim 1$, or usually just $n\,
\delta V \deltaptwo \lesssim 1$, since we will often have
$\deltaptwo \gtrsim 1$.

If two galaxies fall within the same radius-$r_p$ sphere, then
we call this a pair.  If $\delta P$ [cf., equation
(\ref{eqn:twogalaxies})] is the probability to find two galaxies
in a volume $\delta V$, and if $\delta P \ll 1$, then the
spatial density $n_2$ of pairs is $\delta P/\delta V =  n^2
(\delta V)\left(1+\deltaptwo \right)$.  The pair-pair autocorrelation
function $X(r)$, the excess probability over random to find a
pair in each of two volumes $\delta V_1$ and $\delta V_3$
separated by a distance $r_{13}$, is defined by
\begin{eqnarray}
     \delta P &=& n_2^2\, \delta V_1\, \delta V_3 \left[ 1 +
     X(r_{13}) \right] \nonumber \\
     & =& n^4 (\delta V_1)^2 (\delta V_3)^2 \left(1+\deltaptwo \right)^2
     \left[ 1 + X(r_{13}) \right],
\label{eqn:Xdefinition}
\end{eqnarray}
where $\delta P$ is here the joint probability to find one pair in
$\delta V_1$ and another in $\delta V_3$.

A pair of pairs is a quadruplet.  To describe the clustering of
pairs of galaxies, we will therefore need
the four-point correlation function.  The joint differential
probability to find objects in differential volume elements
$\deriv V_1$, $\deriv V_2$, $\deriv V_3$, and $\deriv V_4$
located, respectively, at
positions $\vec r_1$, $\vec r_2$, $\vec r_3$, and $\vec r_4$, is
\citep{Peeblesbook}
\begin{eqnarray}
     \delta P &=& n^4\, \deriv V_1\, \deriv V_2\, \deriv V_3\,
     \deriv V_4 \nonumber \\
     &\times & \left[ 1 + \xi_{12} + \, \, 5\, {\rm
     permutations} \nonumber  \right.  \\
     & + & \zeta(\vec r_1,\vec r_2, \vec r_3) + \, \, 3\, {\rm
     permutations} \nonumber \\
     &+& \xi_{12} \xi_{34} + \, \, 2\, {\rm permutations}
     \nonumber \\
     &+& \left. \eta(\vec r_1,\vec r_2, \vec r_3, \vec r_4) \right].
\label{eqn:fourpoint}
\end{eqnarray}
Here, $\zeta(\vec r_1,\vec r_2,\vec r_3)$ is the reduced (or
``connected'') three-point correlation function, $\eta$ is the reduced
(connected) four-point correlation function, and we have introduced
the shorthands $r_{ij} \equiv |\vec r_i - \vec r_j|$ and also
$\xi_{ij} \equiv \xi(|\vec r_i-\vec r_j|)$.  The quantity in brackets
is the complete four-point autocorrelation function.  For Gaussian
perturbations, $\zeta=\eta=0$.

To find the pair autocorrelation function, we now consider the
case where two of the galaxies (1 and 2) are in one volume
($\delta V_1$) centered at $\vec r_1$ and the other two (3 and
4) are in another ($\delta V_3$) centered at $\vec r_3$.  We
also assume that the separation $| \vec r_1 -\vec r_3| \gg
r_p$.  The joint probability to find two galaxies in $\delta V_1$ and
two in $\delta V_3$ is thus
\begin{eqnarray}
     \delta P &=& n^4 \int_{\delta V_1} \deriv^3 x_1  \int_{\delta V_1}
      \deriv^3 x_2  \int_{\delta V_3} \deriv^3 x_3  \int_{\delta
      V_3} \deriv^3 x_4 \nonumber \\ 
      &\times& \left[ 1+ \xi_{12} +\xi_{34}
      + 4 \xi_{13} +\xi_{12}\xi_{34} \nonumber \right. \\
      &+& 2 \xi_{13}^2 + 2\zeta(r_{12},r_{13},r_{13}) +
      2\zeta(r_{34},r_{13},r_{13}) \nonumber \\
      &+& \left. \eta(r_{12},r_{13},r_{14},r_{23},r_{24},r_{34})
      \right];
\label{eqn:fourpointreduced}
\end{eqnarray}
note that in this equation (only), $\xi_{12} = \xi(|\vec x_1
-\vec x_2|)$ and similarly for $\xi_{34}$.  We next note that
\begin{eqnarray}
     \int_{\delta V_1} \deriv^3 x_1 \int_{\delta V_1} \deriv^3 x_2
     \int_{\delta V_3} \deriv^3 x_3\zeta (\vec x_1,
     \vec x_2,\vec x_3) \nonumber \\
     = (\delta V)^3\VEV{\delta_p^2(\vec r_1) \delta_p(\vec r_3)}_c,
\end{eqnarray}
the (reduced) three-point correlation function (with two of the
three points coincident) for the smoothed density field, and
\begin{eqnarray}
     \int_{\delta V_1} & \deriv^3 x_1 & \int_{\delta V_1} \deriv^3 x_2
     \int_{\delta V_3} \deriv^3 x_3 \int_{\delta V_3} \deriv^3 x_4 \nonumber
     \\
     & & \times\eta (\vec x_1,\vec
     x_2,\vec x_3,\vec x_4)\nonumber  \\
     & = &
     (\delta V)^4\VEV{\delta_p^2(\vec r_1) \delta_p^2(\vec r_3)}_c,
\end{eqnarray}
a (reduced) four-point correlation function.  Equating equations
(\ref{eqn:Xdefinition}) and (\ref{eqn:fourpointreduced}), we find
\begin{eqnarray}
 X(r) &=& \left[ 4  \xi(r) + 2 \xi^2(r) + 4 \VEV{\delta_p^2(\vec
     x)\delta_p(\vec x +\vec r)}_c \right.  \nonumber \\
      & & + \left. \VEV{\delta_p^2(\vec x) \delta_p^2(\vec x+\vec
     r)}_c \right]/ \left(1+\deltaptwo\right)^2.
\label{eqn:paircorrelation}
\end{eqnarray}
This result becomes exact in the limit that $r \gg r_p$ and
$n\delta V \left(1+\deltaptwo\right) \ll1$, and it is valid
for any galaxy-galaxy two-point, three-point, and four-point
autocorrelation functions.  We thus find that the calculation of
the pair correlation function reduces to the calculation of the
correlation of the density $\delta_p$ with $\delta_p^2$ and the
autocorrelations of $\delta_p^2$, a result that should come as
no surprise.

We will define the effective pair bias via $b_p^2 \equiv
[X(r)/\xi(r)]$; it is the excess bias of pairs relative to individual
objects.  Note then that, in the language of Section 2, the net merger
bias is $b_m = b_h \, b_p$.

\section{Pair clustering for Gaussian perturbations}

For Gaussian perturbations, $\zeta=\eta=0$ and the pair-pair autocorrelation
function simplifies to
\begin{equation}
     X(r) = \frac{4 \xi(r) + 2 [\xi(r)]^2}{\left(1+ \deltaptwo \right)^2}.
\label{eqn:Gpaircorrelation}
\end{equation}
In the limit of weak correlations, $\deltaptwo,\xi \ll 1$, $X(r)
\simeq 4 \xi(r)$.  This is easy to understand: given two galaxies in
the first cell, each contributes a factor $\xi(r)$ to the excess
probability to find a galaxy in the second cell (at least to linear
order), and for $X(r)$ there are two such galaxies in the second cell.
Although of interest academically, this limit is probably not relevant
for galaxies or clusters of galaxies, as a value $\deltaptwo \lesssim
1$ requires that we deal with objects that are so rare that their mean
separations are $\gtrsim$\,Mpc.

If $\xi(r) \lesssim 1$ and $\deltaptwo \gtrsim 1$, then the clustering
of pairs is suppressed relative to that of individual galaxies, a
consequence of the scarcity of pairs relative to individual galaxies.
In the limit of strong clustering, $\xi(r),\deltaptwo \gg 1$, the pair
correlation function becomes $X(r) \simeq 2 [\xi(r)]^2/\deltaptwo^2$,
which is again suppressed relative to the galaxy correlation function.
The applicability of this limit, however, should be questioned, as
$\xi \gtrsim 1$ generally implies non-Gaussian perturbations.
Interestingly, this simple exercise implies that merger bias can
operate in different directions, depending on the regime of
interest---as indeed the simulations discussed above find.

\section{Clustering of Gaussian peaks}

We have just seen that if objects trace the distribution of mass in a
system with Gaussian perturbations with some specified correlation
function, then the pair correlation function is suppressed relative to
the normal correlation function, unless the correlations are weak, in
which case it can be enhanced by up to a factor of 4.  If, however,
objects form only at high-density peaks of a primordial density
distribution, then the distribution of these objects will be
non-Gaussian.  That this is true is easy to see.  The one-point
probability distribution function for Gaussian perturbations is
$P(\delta) \propto e^{-\delta^2/2\sigma^2}$, where $\sigma^2$ is the
variance.  This distribution has zero mean, no skewness, no kurtosis,
and no higher-order (reduced) cumulants.  The one-point probability
distribution of high-density peaks is $P(\delta)\propto
e^{-\delta^2/2\sigma^2}$ for $\delta>\nu\sigma$ and $P(\delta)=0$ for
$\delta<\nu\sigma$.  This distribution has nonzero mean, nonzero
skewness, kurtosis, etc.

This non-Gaussianity introduces non-zero reduced three-point and
four-point correlation functions
\citep{PolitzerWise,BBKS,JensenSzalay,MelottFry}, even if the
total-density--perturbation amplitude is linear, $\xi\lesssim1$.
Although the exact expressions can be complicated, they simplify
considerably when $\nu \gg 1$.  In this limit, the full $n$-point
correlation function can be written in terms of the galaxy two-point
correlation function $\xi_g(r)$ as \citep{PolitzerWise}
\begin{equation}
1 + \xi_g^{(n)}(\vec r_1,...,\vec r_n) \simeq \prod_{i>j} [
    \xi_g(r_{ij}) + 1 ].
\label{eqn:peaknpt}
\end{equation}
The galaxy correlation function $\xi_g(r)$ is the correlation function
for the objects, rather than for the total mass.  Thus, we can have
$\xi_g \gtrsim1$ even in the linear regime, $\xi(r) \lesssim1$, if the
objects are highly biased tracers of the mass distribution.  In this
case, we can simply replace the expression in brackets in equation
(\ref{eqn:fourpointreduced}) with $[1+\xi_g^{(4)}]$ from equation (\ref{eqn:peaknpt}).  Then, the pair
autocorrelation function becomes
\begin{equation}
X(r) \simeq [1+ \xi_g(r)]^4 -1.
\label{eqn:centralresult}
\end{equation}
This equation is the central result of this Section.  It says that,
if objects trace the distribution of peaks in a Gaussian density
distribution, then the clustering of pairs can be strongly enhanced
relative to the clustering of individual objects.  
Equation~(\ref{eqn:centralresult}) is valid for highly-biased objects ($\nu \gg 1$) on scales at which the underlying matter fluctuations are linear (even if fluctuations in the population of the objects is \emph{not} small; \citealt{PolitzerWise}).  It thus applies to haloes well above the characteristic mass scale (such as submillimetre galaxies at $z \sim 3$ or extremely massive clusters at the present day).
Physically,
higher-order clustering---in particular, the four-point correlation
function from equation (\ref{eqn:peaknpt}), which provides nonzero
reduced three- and four-point functions---of high-density peaks is
enhanced with this type of non-Gaussianity, and this favors the
clustering of pairs over individual objects.  Thus, if mergers can be
equated with close pairs of galaxies, we do expect a significant
merger bias in the limit $\nu \gg 1$.

\section{Quasilinear perturbations}

Equation (\ref{eqn:paircorrelation}) shows that the pair correlation
depends on the three- and four-point correlation functions.  The
previous Section showed that such terms do appear if galaxies are
associated with peaks in the density field.  However, another way to
produce non-zero higher-order correlations is through gravitational
processes, and it is interesting to consider how such processes could
affect pair correlations (and hence the merger bias).  We therefore
next consider objects that are distributed like the mass for a
non-Gaussian mass distribution produced by gravitational
amplification, to the quasilinear regime, of primordial Gaussian
perturbations.  At redshift $z=0$, the quasilinear regime occurs at
$\sim10$ Mpc; at redshift $z=3$, it occurs at $\sim 1$ Mpc.  The
bispectrum and trispectrum for this case can be calculated from
cosmological perturbation theory and from them the three- and
four-point correlation functions.  The expressions can be quite
formidable \citep{Goroff}, but fortunately for us, Bernardeau (1996;
see also \citealt{Roman}) has calculated the quantities required here.
In particular, in the nonlinear regime,
\begin{equation}
     \VEV{\delta_p^2(\vec x_1) \delta_p(\vec x_2)}_c = 
     C_{2,1} \VEV{ \delta_p^2} \xi(|\vec x_1 -\vec x_2|),
\end{equation}
and
\begin{equation}
     \VEV{\delta_p^2(\vec x_1) \delta_p^2(\vec x_2)}_c = 
     C_{2,1}^2 \VEV{ \delta_p^2}^2 \xi(|\vec x_1 -\vec x_2|),
\end{equation}
where
\begin{equation}
     C_{2,1} = \frac{68}{21} + \frac{1}{3} \frac{ \deriv \log
     \deltaptwo}{\deriv \log r_p}.
\end{equation}
In the limit that $\VEV{\delta_p^2} \gg 1,\xi$, we find
\begin{equation}
     X(r) \simeq C_{2,1}^2 \, \xi(r).
\label{eqn:quasiresult}
\end{equation}
We note that $\deriv \log \left(\deltaptwo \right)/\deriv \log r_p =
\deriv \log \xi/\deriv \log r$.  For the scales probed by Lyman-break
galaxies, the linear-theory correlation function is roughly $\xi
\propto r^{-2}$, while stable clustering leads to a correlation
function $\xi(r) \propto r^{-1.8}$.  For these correlation-function
scalings, $X(r) \simeq 7\, \xi(r)$; i.e., pairs are biased by roughly
a factor of 2.6 relative to galaxies.  If, on the other hand, $\xi(r)
\propto$\,constant at small radii (as expected for $P(k) \propto k^n$
with $n=-3$), then $X(r) \simeq 10\, \xi(r)$.  We thus find that in
the quasilinear regime, pairs can be biased, perhaps strongly so,
compared with the individual objects, even if they trace the mass.
This could further enhance the clustering of mergers, if they are
associated with pairs of objects.  
We emphasize that equation~(\ref{eqn:quasiresult}) is applicable on scales at which the underlying mass perturbations have $\xi \sim 1$ and assumes that the objects of interest exactly trace the mass distribution.  They are thus only directly applicable in the limited regime of relatively unbiased objects on moderately small scales, although the qualitative results likely apply to more biased objects as well (see the discussion at the end of Section 7).

\section{Halo clustering model}

We will now briefly consider pair clustering in the highly nonlinear
regime.  In this case, perturbation theory is no longer appropriate,
so we will turn to the halo model of the density field.
The halo clustering model postulates a distribution of virialized
dark-matter haloes, each with a radial ($r$) density profile
$\rho_h(m;r)$ that depends on its mass $m$.  On large scales, the
clustering is that of biased peaks, possibly in the quasilinear
regime, which we already considered above.  On nonlinear scales, the
clustering is described within individual haloes.
Of course, in this ``one-halo" regime, the distribution of objects is ultimately due to the interactions between them (such as dynamical friction acting on satellite galaxies).  Our treatment is thus only approximate:  it predicts the clustering of pairs given a density profile and implicitly ignores interactions.  It could, nevertheless, be useful inside clusters of galaxies in which a population of small ``tracer" haloes orbit in a potential dominated by the massive cluster.  

For the purposes of illustration, we suppose that all haloes have
the same mass and power-law radial density
profile: $\rho \propto r^{-\gamma}$ for $r<R$, and $\rho(r) = 0$
for $r>R$.  We will only consider correlations on small scales,
within an individual halo (which should be appropriate on small scales in
the highly nonlinear regime).  The autocorrelation function for
the mass is then \citep{Scherrer,CooraySheth},
\begin{equation}
     \xi(|\vec r_1-\vec r_2|) = \frac{ \VEV{\rho(\vec
     r_1)\rho(\vec r_2)}}{ \VEV{\rho}^2} -1,
\end{equation}
where the angle brackets denote an average over all space.  The
mean density is $\VEV{\rho} = n_{\rm halo} M$, where $n_{\rm
halo}$ is the spatial number density of halos and $M$ is the
halo mass, and
\begin{equation}
      \VEV{\rho(\vec r_1)\rho(\vec r_2)} = n_{\rm halo} \int \deriv^3 x\,
     \rho(|\vec r_1 - \vec x|) \rho( |\vec r_2 - \vec x|).
\label{eqn:haloauto}
\end{equation}
The integral in
equation~(\ref{eqn:haloauto}) is particularly simple at zero
lag, where the autocorrelation function for the mass is
\begin{equation}
     \xi(r=0) = (4 \pi n_{\rm halo} R^3)^{-1} \frac{
     (3-\gamma)^2}{(3-2\gamma)} -1,
\end{equation}
for $\gamma<3/2$.  For $\gamma>3/2$, the divergence at the
$r\rightarrow0$ limit of the integrand can be tempered by measuring correlations over a finite smoothing volume of radius $r_s$ (as would occur in any physical observation).  Thus, for $\gamma>3/2$, we find
\begin{equation}
     \xi(r=0) = (4 \pi n_{\rm halo} R^3)^{-1} \frac{
     (3-\gamma)^2}{|3-2\gamma|} \left(\frac{r_s}{R}
     \right)^{3-2\gamma} -1. 
\end{equation}
For $r\lesssim R$ and $\gamma>3/2$ (and for $n_{\rm halo} R^3
\ll 1$), the mass correlation function scales with radius $r$ as
$\xi(r) \propto r^{3-2\gamma}$; for $\gamma<3/2$, it decreases
less rapidly with radius.  For $\gamma=3/2$, the power laws
are replaced by logarithms.

The pair correlation function follows simply by noting that
pairs are distributed in the halo as $\rho^2$.  We can therefore
simply replace $\gamma\rightarrow 2\gamma$ in the results for
the mass correlation functions.  Thus, for $\gamma<3/4$, the
zero-lag pair correlation function is
\begin{equation}
     X(r=0) = (4 \pi n_{\rm halo} R^3)^{-1} \frac{
     (3-2\gamma)^2}{(3-4\gamma)} -1,
\end{equation}
and for $\gamma>3/4$,
\begin{equation}
     X(r=0) = (4 \pi n_{\rm halo} R^3)^{-1} \frac{
     (3-2\gamma)^2}{|3-4\gamma|} \left(\frac{r_s}{R}
     \right)^{3-4\gamma} -1. 
\end{equation}
The pair correlation function scales, for $\gamma>3/4$, with
radius $r$ as $X(r)\propto r^{3-4\gamma}$, and it decreases less
rapidly with $r$ for $\gamma<3/4$.
For $3/4 <\gamma <3/2$, the pair correlation diverges (modulo
the smoothing) at small radii, while the mass correlation
function approaches a constant as $r\rightarrow 0$.

We thus see that the distribution of pairs and mass differ, and
thus that there should be a (scale-dependent) bias between
them.  Our calculation is applicable in the nonlinear regime,
when the correlation function is measured at distances $r \ll
R$.  The pair bias can then be approximated by the square root of the
ratio of zero-lag biases.  For example, if $\gamma=1/2$, then the
pair bias evaluates to $b_p = 2^{5/2}/5\simeq 1.1$.  
For $\gamma \rightarrow0$, the pair bias approaches 1, which is
what we expect for objects distributed uniformly in a halo.  The zero-lag
bias may be considerably larger for $3/4<\gamma<3/2$, when the
pair correlation function diverges as $r\rightarrow0$, while the
mass correlation does not.

So far, we have considered pair correlations for a highly biased
population in the linear regime as well as for a population that
traces the mass in the quasilinear regime and in the nonlinear regime.
What about pair correlations for a highly biased population in the
quasilinear or nonlinear regimes?  It has been argued that in the
quasilinear regime, highly-biased tracers are more likely to be found
in denser regions \citep{ColeKaiser,ShethTormen}; calculation of the
pair correlation for a population biased in Lagrangian space evolved
into the quasilinear regime could be done following the techniques of
\citet{Fry}, \citet{Catelanetal}, and \cite{Paolo}, but we leave that
for future work.  And what about the nonlinear regime?  Numerical
simulations have suggested that the distribution of primordial density
peaks in larger virialized haloes (i.e., the nonlinear regime) is more
highly peaked toward the centers than the mass as a whole
\citep{Santos,Moore98,WhiteSpringel,Diemand}.  If so, and if, as we
have seen, the bias of pair correlations is enhanced with steeper
density profiles, then the bias of pair correlations for rare objects
in the quasilinear and nonlinear regimes may be even further enhanced.

\section{Discussion}

In this paper, we have investigated the implications of the
extended Press-Schechter and \citet{Mo96} biasing scheme for merger bias
and pointed out some shortcomings and ambiguities in this approach.
In particular, we showed that this approach yields \emph{no} merger
bias, but only because it explicitly ignores the variation of merger
rates with the large-scale density field.  We then showed that a
simple model in which the merger rate scales only with the halo
abundances predicts that mergers of massive galaxies will be more biased
than the halo population but that mergers of small galaxies will be
less biased.  Furthermore, the merger bias will evolve significantly
with redshift.  These may provide useful clues to reconciling the
various simulations \citep{Evan,percivaletal,Gaoetal}.  However, these
techniques are clearly inadequate for understanding merger bias on any
quantitative level (at least until a self-consistent merger kernel is
available).

We therefore moved on to hypothesize that close pairs in a clustering
model are likely to yield mergers.  We thus studied the clustering of
close pairs in a variety of models in which objects Poisson sample (1)
the mass in a Gaussian random field; (2) the high-density peaks in a
Gaussian random field; (3) the mass in the quasilinear regime; and (4)
the mass in virialized haloes with power-law density profiles.  We
find that in many (though not all) cases, close pairs can be more
highly clustered than individual objects.  If so, and if close pairs
are likely to lead to mergers, then the clustering of objects that
have undergone recent mergers can be enhanced relative to the
clustering of individual haloes of comparable masses.  We have thus
shown that, in the simplest picture of mergers, an extra bias (of some
magnitude) is generic to most clustering models.  The actual magnitude
of the bias (or the lack of it, as in the simulations of
\citealt{percivaletal,Gaoetal}) is therefore revealing something
fundamental about the halo-merging process---an area in need of
substantial theoretical insight \citep{Benson05}.

Even if we do identify close pairs with mergers, there are still a
multitude of theoretical steps---each fraught with considerable
uncertainties---that must be taken to connect close pairs of galactic
haloes with, e.g., the observational constraints on LBGs.  
We have considered the behavior under a variety of limits, but the
more general case must be treated numerically.  Still, it is
interesting to investigate whether pair biasing might be in the right
ballpark to account for the discrepancy between the LBG dynamical and
clustering masses.  According to \citet{Adelberger98}, the bias of
LBGs is $b_{\rm LBG} \sim 4.0$, roughly consistent with that expected for
$\sim10^{12}\, {\rm M}_\odot$ objects (see also \citealt{Adelberger05}, who
estimate a similar median mass for a larger sample of objects at
$z=3$).  Although the abundance of haloes with such masses is
consistent with the abundance of LBGs, it requires that {\it every}
such halo house a galaxy that produces stars at a prodigious rate
\citep{AdSteid}.  On the other hand, the linewidths and kinematics of
LBGs suggest masses closer to $\sim10^{11}\, {\rm M}_\odot$
\citep{Pettini01,Erb03}.  Haloes of these masses have a much higher
abundance, allowing consistency with the LBG abundance if the
efficiency for $\sim10^{11}\, {\rm M}_\odot$ haloes to produce extremely
luminous objects is relatively low, $\sim10\%$---understandable,
perhaps, if only recent mergers of $\sim10^{11}\, {\rm M}_\odot$ haloes
produce LBGs.  (An alternate possibility is that dynamical mass measurements are only sensitive to a small fraction of the halo and that LBGs are ubiquitous in large dark matter haloes; \citealt{Cooray05}.)

The only remaining problem with the small-mass LBG scenario is why the bias $b_{\rm LBG} \sim 4$ is so much
larger than the bias $b_{11}\approx 2.4$ expected for a sample with
$\sim10^{11}\, {\rm M}_\odot$ haloes.  \citet{Adelberger98} measure the
clustering through a counts-in-cells analysis within boxes of size $11.4\,
h_{100}^{-1}$ Mpc.  This is within the linear regime at redshifts
$z\sim3$, and with an expected bias $b_{11}\approx 2.4$, the variance in
the $\sim10^{11}\,{\rm M}_\odot$ halo distribution is $\sigma_{\rm gal}
\simeq 0.8$.  It is also reasonable to assume a pair spacing with
$\deltaptwo \gg1$.  Although the pair bias implied by
equation~(\ref{eqn:centralresult}) is not linear, most of the weight
for the counts-in-cells analysis occurs at the largest radii.  We thus
estimate from equation~(\ref{eqn:centralresult}) a pair bias (i.e.,
the extra biasing of mergers relative to the objects themselves) of
$b_p = \sqrt{X(r)/\xi_g(r)} \simeq 3.4$.  This is more than enough to
make the net merger bias ($b_m = b_p b_h$) comparable to $b_{\rm LBG}$.
However, note that $\nu \approx 1.6$ for $10^{11}\,{\rm M}_\odot$ haloes, so
the true amplification should be smaller than the $\nu \gg 1$ limit we
have taken.  This may be further augmented by quasilinear effects,
which could contribute a comparable pair bias over some fraction of
the cell.

A similar, though perhaps even more desperate, problem occurs for submillimetre-selected galaxies.  \citet{Blain04} claim that  the clustering of these galaxies indicates halo masses of $\sim 10^{13} \, {\rm M}_\odot$ while kinematic measurements yield values an order of magnitude smaller, even allowing for the mass in the outer regions of the halo.  Our results may help resolve this discrepancy as well, if submillimetre galaxies are the products of recent mergers.  Moreover, \citet{Blain04} measured clustering through the rate of incidence of close pairs in their survey fields.  They assumed a correlation function of fixed shape $\xi_g(r) \propto r^{-1.8}$ and varied its amplitude until they recovered the observed number of pairs; the inferred correlation length could then be matched to a halo mass.  We have shown that the clustering of pairs is \emph{not} the same as the clustering of the underlying objects and depends on the underlying halo population, the scales of interest, and even the relation of haloes to the underlying density field.  The effective pair bias can be significantly larger than the bias of the haloes themselves, so pair-counting techniques must be approached with care.  The precise effects are difficult to predict given the ``pencil-beam" geometries of their surveys, but they certainly merit further study.

Before closing, we note that our results may be applicable
elsewhere as well.  For example, galaxy clusters are highly
biased tracers of the mass distribution today \citep{Bahcall03}.  
Their correlation length may be as large as $\sim25\, h_{100}^{-1}$
Mpc, as opposed to a correlation length $\sim 5-7\, h_{100}^{-1}$ Mpc
for the mass.  If this bias occurs because clusters form at peaks of
the primordial density distribution, then they should experience
higher-order clustering as described in Section 5.  Moreover, at
distances $\gtrsim10\, h_{100}^{-1}$ Mpc, quasilinear effects should
be small.  There will thus be testable predictions for the clustering
of close pairs of clusters, or---if pairs are associated with
mergers---for the clustering of recently merged clusters.  
As another example, non-trivial merger bias would modify the
interpretation of AGN clustering (provided that they are fueled by
merger activity).  This would be particularly important for
understanding their host properties and their lifetimes
\citep{LaFranca98,Haiman01,Martini01,Adelberger05a,Adelberger05b}.  We
leave further discussion of these possibilities to future work.

We thank the referee, L. Miller, for helpful comments.  This work was supported in part by DoE DE-FG03-92-ER40701 and
NASA NNG05GF69G.


\end{document}